\documentclass[pra,aps,amsmath,amssymb,twocolumn,superscriptaddress]{revtex4} \usepackage{times}
\usepackage{hyperref}
\usepackage{amsmath,amssymb}
\usepackage[usenames]{color}
\usepackage{grffile}
\usepackage[pdftex]{graphicx}
\usepackage{amsmath, amstext, amssymb, amsfonts, amsxtra}
\usepackage{textcomp}
\usepackage{xspace}
\DeclareSymbolFontAlphabet{\amsmathbb}{AMSb} 
\usepackage{soul}
\usepackage{xcolor}
\usepackage{epstopdf}
\usepackage{braket}
\usepackage{float}
\usepackage{tikz}
\usetikzlibrary{bending,arrows.meta,fadings}
\usepackage{pgfplots}
\usepgfplotslibrary{statistics}
\usepackage{enumitem}

\setlist{nosep}
\usetikzlibrary{arrows}

\newcommand{\sop}{\sigma}

\newcommand{\smb}[1]{\left(#1\right)}

\newcommand{\pdt}[1]{\frac{\partial #1}{\partial t}}
\newcommand{\cmute}[2]{\left[{#1},{#2}\right]}
\newcommand{\ii}{\mathrm{i}}
\newcommand{\expe}{\mathrm{e}}

\newcommand{\Hs}{H_\mathrm{S}}

\newcommand{\drme}[2]{\mathcal{R}^{#1}\left[#2\right]}

\newcommand{\lux}{Complex Systems and Statistical Mechanics, Physics and Materials Science Research Unit, University of Luxembourg, L-1511 Luxembourg, Luxembourg}
\newcommand{\ieu}{Zhengzhou Information Science and Technology Institute, Zhengzhou 450004, China}
\newcommand{\sutd}{EPD Pillar, Singapore University of Technology and Design, 8 Somapah Road, 487372 Singapore} 
\newcommand{\sutdsci}{Science and Math Cluster, Singapore University of Technology and Design, 8 Somapah Road, 487372 Singapore}

\begin{document}

\title{Many-body open quantum systems beyond Lindblad master equations}               

\author{Xiansong Xu} 
\affiliation{\sutdsci}
\author{Juzar Thingna}
\affiliation{\lux}
\author{Chu Guo} 
\affiliation{\ieu}
\author{Dario Poletti}
\affiliation{\sutdsci} 
\affiliation{\sutd}

\begin{abstract} 
  Many-body quantum systems present a rich phenomenology which can be significantly altered when they are in contact with an environment. In order to study such setups, a number of approximations are usually performed, either concerning the system, the environment, or both. A typical approach for large quantum interacting systems is to use master equations which are local, Markovian, and in Lindblad form. Here, we present an implementation of the Redfield master equation using matrix product states and operators. We show that this allows us to explore parameter regimes of the many-body quantum system and the environment which could not be probed with previous approaches based on local Lindblad master equations. We also show the validity of our results by comparing with the numerical exact thermofield-based chain-mapping approach.
\end{abstract}

\maketitle
\section{\label{Sec:intro} Introduction}
In quantum systems, interactions can induce phases of matter with peculiar properties \cite{SachdevBook}. While it is still a very demanding task to understand the ground-state properties of strongly correlated quantum systems, the study of many-body quantum systems in contact with an environment is a much less explored territory. In this case the environment can significantly alter the properties of the system, either suppressing desired properties or enhancing them \cite{DiehlZoller2008, MullerZoller2012}. For example, the environment can induce dephasing in a system, thus forcing it to lose coherence or to alter or suppress its localization properties \cite{FischerAltman2016, LeviGarrahan2016, EverestLevi2016, ZnidaricGoold2016, ZnidaricVarma2016, MedvedyevaZnidaric2016, VanNieuwenburgFischer2017, XuPoletti2018, VakulchykDenisov2017}. On the other hand, a bath, especially if carefully tailored, can be used to favor condensation \cite{DiehlZoller2008, DiehlZoller2010} or exotic phases of matter in the steady state \cite{MullerZoller2012} or for long times \cite{BernierKollath2013, ShiraiMiyashita2016}. The interplay of strong interaction and dissipation has also been shown to result in nontrivial relaxation regimes, from power law \cite{PolettiKollath2012, CaiBarthel2012} to stretched exponentials \cite{LeviGarrahan2016, EverestLevi2016, PolettiKollath2013} and aging \cite{SciollaKollath2015}. For a review on some aspects of many-body open quantum systems one can refer to \cite{Daley2014}.             
The study of such systems is, however, limited by approximations needed to treat the many-body quantum system and to model the environment and its interaction with the system itself. 
          
The difficulty of studying many-body quantum systems (even when isolated from the environment) stems from the fact that a many-body wave function lives in a space which grows exponentially with the system size. Hence, simulation of such systems would be computationally expensive, even for a few tens of sites. Over the years, various numerical methods have been developed to study such systems, from mean-field \cite{Gutzwiller1963, Gutzwiller1964, Gutzwiller1965, FisherFisher1989, RoksharKotliar1991, JakschZoller1998} to dynamical mean-field theory \cite{MetznerVollhardt1989, GeorgesKotliar, GeorgesRozenberg1996} and quantum Monte Carlo \cite{SenatoreMarch1994, Ceperley1995, FoulkesRajagopal2001}. Another family of methods uses tensor networks \cite{Schollwock2005, VerstraeteCirac2008, Schollwock2011, Orus2014}, especially for one-dimensional systems where they are commonly known as matrix product states (MPSs). In this scenario, tensor network algorithms are implemented in different flavors to search for ground states \cite{White1992} and to compute time evolutions \cite{WhiteFeguin2004, Vidal2004, DaleyVidal2004, VerstraeteCirac2004, ZwolakVidal2004}.

For open quantum systems the computational complexity grows further. In fact, density matrices are described in a space which is the square of that of wave functions. Moreover, the environments need to be modeled appropriately for an accurate description of dissipative effects. For weak system-environment coupling, it is possible to derive various master equations under different assumptions \cite{GardinerZoller2000, PetruccioneBreuer2002, DeVegaAlonso2017}. 

Current studies of large many-body open quantum systems mostly rely on master equations in Lindblad form \cite{GoriniSudarshan1976, Lindblad1976} due to its ease of implementation and computation. In addition, to study large systems, further assumptions on the locality of operators used are required in order to remove the time dependence in the dissipator. However, they may not produce physical results even for weak system-environment coupling \cite{Wichterich2007, Purkayastha2016, XuThingnaWang2017, LevyKosloff2014}, and this is motivating recent research \cite{RivasPlenio2010, TrushechkinVolovich2016, GonzalezAdesso2017, HoferBrunner2017, Rivas2017, MitchisonPlenio2018, WerlangValente2014}. To go beyond the local system operator assumption, one could opt for master equations with a global system operator. Unfortunately, however, these master equations usually work in the eigenbasis where the full energy spectra are required, making it difficult to simulate large quantum systems. Hence, it has not been shown how to simulate large many-body quantum systems with master equations that go beyond the local Lindblad approach. Due to these constraints, a large variety of many-body open quantum systems still remain unexplored.   

Here we show how to realize the Redfield master equation RME, which goes beyond the limits of local Lindblad master equations, by using matrix product states and operators to study larger many-body quantum systems. As an application, we consider an XXZ spin chain with its center site coupled to a thermal bath, and we show the system's response to the thermal bath by analyzing the local magnetization and correlation propagation. We also demonstrate that this approach goes beyond various Lindblad master equation approaches and is consistent with the numerical exact thermofield-based chain-mapping approach \cite{deVegaBanuls2015}.

This paper is organized as follows. In Sec. \ref{Sec:frame}, we give a general form of the Redfield master equation that can be studied via matrix product state and briefly discuss other types of quantum master equations. In Sec. \ref{Sec:RMEMPS}, we propose a possible implementation of the Redfield master equation with matrix product states and operators. As a demonstration of the implementation, we study the dynamics of a spin-1/2 Heisenberg XXZ model described in Sec. \ref{Sec:model}. In \ref{Sec:results}, we show the supremacy of the proposed implementation by comparing to the conventional approach as well as Lindblad master equations. We further show the consistency between our implementation and the numerical exact thermofield-based chain-mapping approach described in Appendix \ref{App:tcmps}. Detailed discussions on the numerical errors of these implementations are presented in Appendix \ref{App:error}.

\section{\label{Sec:frame} Framework}
We consider a time-independent total Hamiltonian $H_{\rm tot}$ including both the system and bath 
\begin{align}  
	H_{\rm tot} = H_{\rm S} + H_{\rm B} + S\otimes B, 
\end{align} 
where $H_{\rm S}$ is the Hamiltonian of the system under consideration, $H_{\rm B}$ is the bath Hamiltonian, and the interaction between system and bath is given by $ S\!\otimes \! B$, where $S$ acts on the system while $B$ acts on the bath. Assuming the system-bath coupling to be weak, and that the initial global density matrix of the system and bath $\rho_{\rm tot}(0)$ is in a separable form $\rho_{\rm tot}(0)\approx \rho(0)\otimes \rho_B$ where the reduced density matrix $\rho(0)$ describes the system while $\rho_B$ is a thermal Gibbs state for the bath at temperature $T$, it is possible to derive a master equation for the evolution of $\rho(t)$ given by   
\begin{align}
	\pdt{\rho\smb{t}}=&-{\ii}\cmute{\Hs}{\rho\smb{t}}+\drme{t}{\rho{\smb{t}}},   
	\label{eq:rme}
\end{align}
which is also known as the Redfield master equation (RME) \cite{Redfield1957}. Here the first term on the right-hand side describes the unitary evolution due to the system Hamiltonian. The dissipation due to the bath is described by a time-dependent superoperator   
\begin{align}
	\label{eq:relaxation}
	\drme{t}{~\cdot~}=& \cmute{\mathbb{S}\smb{t}~\cdot~}{S}+\cmute{S}{~\cdot~\mathbb{S}^\dagger\smb{t} }, \\
	\mathbb{S}\smb{t}=&\int^{t}_0 \tilde{S}\smb{-\tau}C\smb{\tau} d\tau,
	\label{eq:transition}
\end{align}  
with $\tilde{S}\smb{\tau}=\expe^{\ii\Hs\tau}S\expe^{-\ii\Hs\tau}$, while the bath correlation function is $C\smb{\tau}={\rm tr}\!\left(e^{\ii H_B \tau} B e^{-\ii H_B \tau} \; B \;\rho_B\right)$.  Note that we work in units such that $J=\hbar=k_B=1$, where $k_B$ is the Boltzmann constant.

To simulate quantum dynamics by using Eq.~(\ref{eq:rme}), one would typically diagonalize the system Hamiltonian $H_{\rm S}$ and express the terms of (\ref{eq:relaxation}) in the energy eigenbasis. Such an approach strongly limits the size of the systems that can be studied.  
For the long time dynamics or steady states, one could evolve the system under a time-independent dissipator $\drme{\infty}{~\cdot~}$ with the transition operator $\mathbb{S}(\infty)$. 
For clarity, we refer to it as the time-independent Redfield master equation (iRME), in contrast to the time-dependent one in Eq. (\ref{eq:rme}). 

In order to investigate larger systems, Lindblad master equations with short range operators are typically used. The advantage of such a master equations is that they can be simulated very effectively with MPS algorithms, either using a trajectory method \cite{MolmerDalibard1992, Daley2014} or the purification of the density matrix \cite{VerstraeteCirac2004}. 
A common microscopic derived Lindblad master equation with local operators relies on the local Hamiltonian approximation and a high-temperature condition \cite{Wichterich2007}, and it is known as the local Lindblad master equation (LLME). In this case, the transition operator $\mathbb{S}(\infty)$ is governed by an approximated local system Hamiltonian (i.e., with intersite coupling terms ignored).     

Another archetypal approximation is to take the singular coupling limit master equation (SCME) \cite{GoriniKossakowski1976, Palmer1977}. In this limit, the correlation function is approximated as $C\left(\tau\right)\approx 2a\delta\left(\tau\right)$, where $a$ depends on the bath model. The corresponding transition operator $\mathbb{S}(\infty)$ then reduces to $aS$ and as a result, the dissipator $\drme{\infty}{~\cdot~}$ becomes local and in Lindblad form too, thus allowing efficient evolution with MPSs. 

\begin{figure}[t]
	\centering
	\includegraphics[width=\columnwidth]{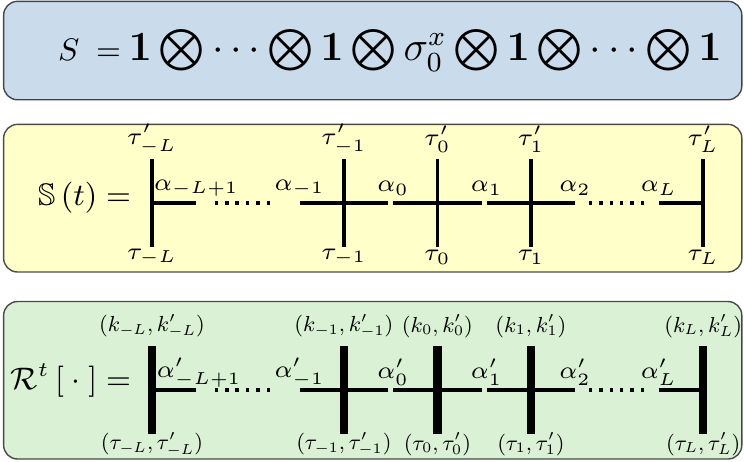}
	\caption{Illustration of the MPO representation of $\mathbb{S}(t)$ and $\mathcal{R}^t$. Starting from the system-bath coupling operator $S$ we evaluate $\mathbb{S}(t)$ via Eq.~(\ref{eq:transition}) and $\mathcal{R}^t$ via Eq.~(\ref{eq:relaxation}). At site $l$ the physical indices of the MPO tensor for $\mathbb{S}$ are $\tau_l$ and $\tau_l'$, while the auxiliary indices are $\alpha_l$ and $\alpha_{l+1}$. For $\mathcal{R}^t$ the tensor at site $l$ has physical indices given by the tuples $(\tau_l, \tau_l')$ for the input and $(\kappa_l, \kappa_l')$ for the output, while the auxiliary indices are $\alpha'_l$ and $\alpha'_{l+1}$. }
	\label{fig:Fig1}
\end{figure}
            
\section{\label{Sec:RMEMPS}Redfield dynamics with matrix product states} 
In order to accurately compute the evolution of a many-body open quantum system, it would be useful to develop a way to compute Eq.~(\ref{eq:rme}) with MPSs, which would allow one to significantly increase the size of the systems currently studied by diagonalizing the system Hamiltonian $H_{\rm S}$. In the following we explain how this can be done. It is possible to describe wave functions and density matrices, even exactly, as a product of tensors \cite{Schollwock2011} with three indices, one for the physical dimension (e.g., of the size of the local Hilbert space), and two auxiliary dimensions (of a maximum size called the bond dimension $D$). Operators acting on a state can be described by linear maps from MPS to MPS, which are called matrix product operators (MPOs). An MPO is a tensor with four indices, one for the input and one for the output physical dimensions, and two auxiliary dimensions of maximum size $D_W$ (the MPO bond dimension).            
We first rewrite the system density matrix $\rho(t)$ as an MPS \cite{VerstraeteCirac2004} and the operators acting on it as MPOs. The MPO representing $S$ is then evolved in time to obtain $\tilde{S}(\tau)$ using a Trotter decomposition at second order. The convolution in Eq.~(\ref{eq:transition}) to compute $\mathbb{S}$ is evaluated subsequently using Romberg integration.          
The algorithm to evaluate $\mathcal{R}^t$ is described pictorially in Fig.~\ref{fig:Fig1}. 
After having obtained $\mathcal{R}^t$ we can use the Runge-Kutta method to evolve $\rho(t)$ using Eq.~(\ref{eq:rme}) \cite{details}.   
We should here comment on the simulability of the evolved and convoluted MPO for the time evolution of the density operator represented by an MPS. In general, for time evolution one can either evolve the state, the operators, or a mixture of both. In practice, when using tensor networks, the best approach depends on the system studied. For instance, an evolution in the Heisenberg picture can be chosen both for isolated \cite{MullerHermesBanuls2012, MuthFleischhauer2011} and open systems \cite{HartmannPlenio2009, PizornTroyer2014, KarraschMoore2012}. However, in general, the time evolution of an operator may require an exponentially increasing amount of memory. In our case, and for the times considered, the decay of the correlations in the bath helps in representing accurately the evolution dynamics of the system while using MPOs of manageable size. For more details on the convergence of numerical simulations, see Appendix \ref{App:rmemps-error}.


\section{\label{Sec:model} Model} 
The methods described above could be applied to a broad range of physical systems. Here we consider a spin-$1/2$ Heisenberg XXZ spin chain with $(2L+1)$ sites, with         
\begin{align}
	\Hs =\!\!\sum_{l=-L}^{L-1}\left[J\left(\sop^{x}_l \sop^x_{l+1}+\sop^y_l \sop^y_{l+1}\right)+\Delta\sop^z_l\sop^z_{l+1}\right]+h\!\!\sum_{l=-L}^{L}\sop^z_l, \nonumber
\end{align}
where $h$ is a uniform magnetic field, and the elements of $\sop_l^\alpha$ are given by the Pauli matrices for $\alpha=x,\;y$, or $z$. $J$ and $\Delta$ denote the tunneling strength and interaction strength, respectively \cite{hfield}.  
The central site ($l=0$) of the spin chain is coupled to a harmonic oscillator bath, with bath Hamiltonian $H_B=\sum^\infty_{n=1}\left[\frac{p_n^2}{2m_n}+\frac{m_n\omega_n^2 x_n^2}{2}\right]$, through the system operator $S=\sop^x_0$ and $B=-\sum^\infty_{n=1}c_n x_n$, where $c_n$ is the system-bath coupling constant for the $n-$th mode.  
The bath properties can be characterized by the spectral function ${\rm J}(\omega)$
\cite{spectral}. In the following, we consider an Ohmic bath with an exponential cutoff, i.e., ${\rm J}(\omega)=\gamma \omega\exp{(-\omega/\omega_{\rm c})}$ \cite{DeVegaAlonso2017, spectral}, and where $\gamma (\propto\sum_n c_n^2)$ is the dissipation strength. It also follows that in the singular coupling limit the prefactor $a=\gamma T$.    
We consider a system with 21 sites, i.e., $L=10$, which cannot be simulated via conventional Redfield master equation approaches ($L\approx 4$, that is, 9-10 sites at most). As an initial condition we choose a fully polarized initial state $\ket{\Psi_0}=\ket{\downarrow\downarrow\downarrow\cdots\downarrow\downarrow\downarrow}$, which is an eigenstate of the system Hamiltonian and it evolves only due to the coupling to the bath.

\begin{figure}[t] 
    \includegraphics[width=\columnwidth]{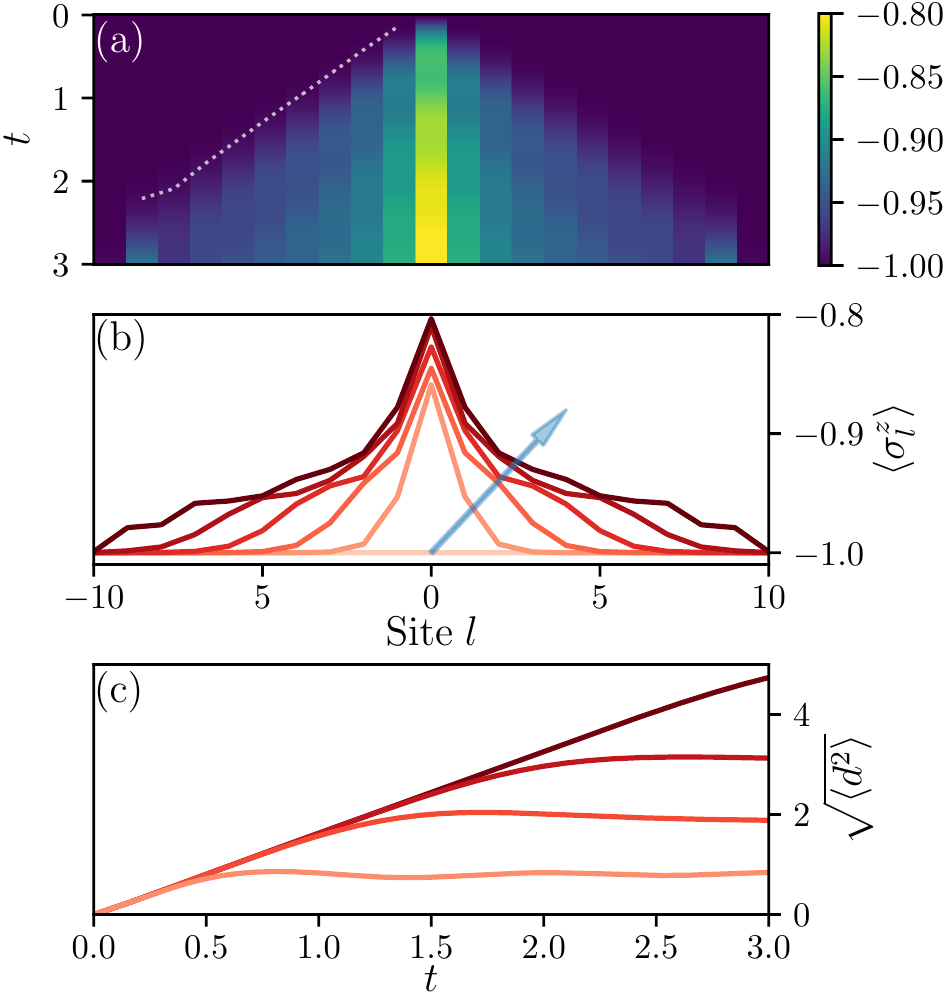}
    \caption{(a) Contour plot of $\langle \sigma^z_l \rangle$ with Redfield master equation (\ref{eq:rme}). (b) Local magnetization profile for times $t=0$, $0.5$, $1$, $1.5$, $2$, $2.5$ (darker lines for larger times). (c) $\sqrt{\braket{d^2}}$ as a function of time from Eq.~(\ref{eq:rme}) (red solid lines), for sizes $5$, $9$, $13$, and $21$ (darker lines for larger systems). Other parameters: $\Delta=5$, $h=0.5$, $\omega_{\rm c}=20$, $T=2$, $\gamma=0.02$. }
	\label{fig:fig2}
\end{figure}

\section{\label{Sec:results} Results}
In Fig.~\ref{fig:fig2}(a) we show the open system dynamics for the fully polarized state $\ket{\Psi_0}$ (the white dotted line depicts a linear propagation). This is expressed more clearly in Fig.~\ref{fig:fig2}(b), which shows cuts, at different times, of panel (a). For a more quantitative analysis we study the variance of the spreading of the magnetization, given by
\begin{align}
	\braket{d^2} = \sum_l \braket{\sop^{u}_l} l^2 \; / \; \sum_l \braket{\sop^{u}_l} , 
\end{align}
where $\sop^{u}_l=\sop^+_l \sop^-_l$ with $\sop^\pm_l=(\sop_l^x\pm{\rm i}\sop_l^y)/2$.
The evolution of $\sqrt{\braket{d^2}}$ is linear due to the fact that an excitation, after it is introduced by the bath, propagates ballistically. For the dissipative evolution we have considered different system sizes so as to show how quickly finite-size effects can play an important role and limit the predictive power.

\begin{figure}[t]
	\includegraphics[width=\columnwidth]{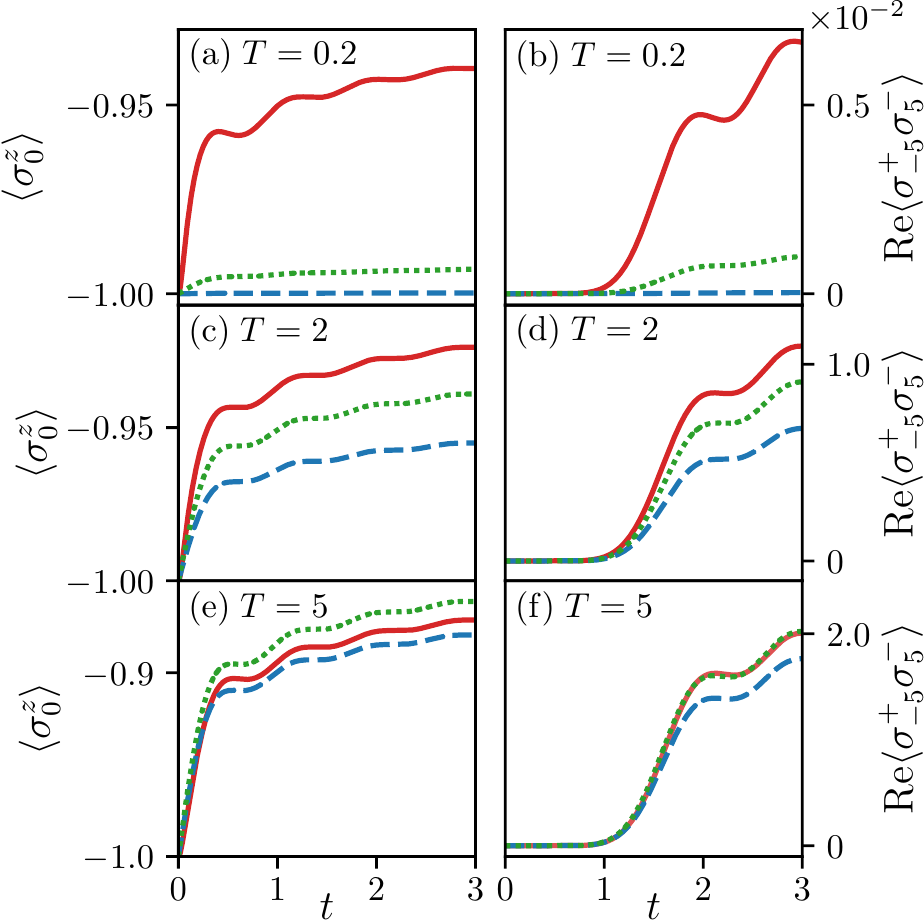}
	\caption{Evolution of local magnetization $\langle\sop^z_0\rangle$, panels (a,c,e) and real part of the long distance correlation $\braket{\sop^+_{-5}\sop^-_{5}}$, panels (b,d,f) as functions of time. The evolutions are computed using RME (red solid lines), SCME (green dotted lines) ,and LLME (blue dashed lines) master equations. Panels (a,b) are for $T=0.2$, (c,d) are for $T=2$, and (e,f) are for $T=5$. Other parameters: $\Delta=0.5$, $h=0.5$, $\omega_{\rm c}=20$, $\gamma=0.02$.}
	\label{fig:fig3}
\end{figure}

\begin{figure}[t]
	\centering
	\includegraphics[width=\columnwidth]{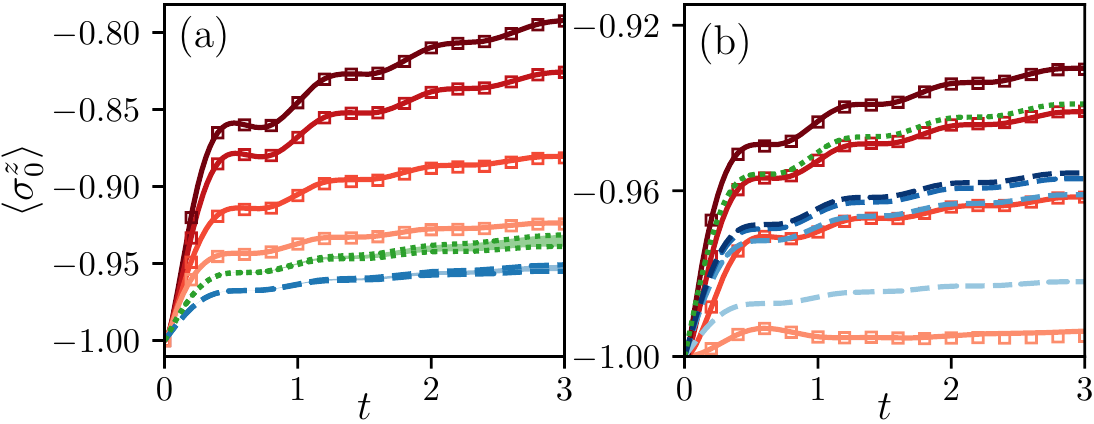}
        \caption{Local magnetization $\braket{\sop^z_{0}}$ vs time $t$ (a) for different interactions $\Delta= 0.5,\; 1.5,\; 3,\; 5$ at $\omega_{\rm c}=20$ or (b) for different cut-off frequencies $\omega_{\rm c}= 1,\; 5,\;10,\;15$ at $\Delta=0.5$, computed from RME (red solid lines with color gradient), SCME (green dotted lines), and LLME (blue dashed lines) for $21$ sites with $\gamma=0.02$. Darker colors imply larger interactions or cut-off frequencies. In all panels, $T=2$. The squares in panels (a) and (b) correspond to the numerically exact TCMPS approach. The parameters used for the TCMPS approach can be referred to Appendix \ref{App:tcmps-error}}.
	\label{fig:fig4}
\end{figure}

We now compare the results of our approach to those of the LLME and SCME. We study two quantities, the local magnetization in the center $\braket{\sop^z_{0}}$, Figs.~\ref{fig:fig3}(a), \ref{fig:fig3}(c), and \ref{fig:fig3}(e), and the correlation between two distant sites $\braket{\sop^+_{-5}\sop^-_{5}}$, Figs.~\ref{fig:fig3}(b), \ref{fig:fig3}(d), and \ref{fig:fig3}(f), for different bath temperatures $T$. For low temperatures, Fig.~\ref{fig:fig3}(a) and \ref{fig:fig3}(b), the dynamics of the Lindblad master equations (dashed blue line for LLME and green dotted line for SCME) is much slower than the more accurate RME (red continuous line). In fact, the derivation of both LLME and SCME requires a high-temperature approximation. As $T$ increases, the curves approach each other, but even for $T=5$, while the evolution is similar, the difference between the various Lindblad master equations and RME is sizable. 

It is important to probe the performance of these master equations for varying many-body interaction strength $\Delta$. In Fig.~\ref{fig:fig4}(a) we show the local magnetization $\langle \sigma_0^z \rangle$ versus time as we vary $\Delta$. We observe that the Redfield dynamics is strongly affected by $\Delta$ (red continuous lines from light to dark as $\Delta$ increases); however, the evolution of both Lindblad master equations (green dotted curves for SCME and blue dashed lines for LLME) does not vary significantly with $\Delta$ but changes only in the shaded regions. This implies that these Lindblad master equations are unable to accurately capture the effect of strong interaction, effectively approximating the many-body physics in this system. 

We also study the effect of bath cutoff frequency $\omega_{\rm c}$, which modifies how different energy levels are coupled to the bath. In Fig.~\ref{fig:fig4}(b) we show $\langle \sigma_0^z \rangle$ as a function of $t$ for various cut-off frequencies $\omega_{\rm c}$. The SCME cannot probe the differences in $\omega_{\rm c}$, and in fact, there is a single green dotted line. The LLME can vary with $\omega_{\rm c}$, but it is not accurately reproducing the RME, even in the weak interaction regime. In particular, even for a highly Markovian environment (i.e., dark red), the LLME shows a strong deviation from RME (see Appendix \ref{App:rme} for more details).  

We benchmark the Redfield dynamics with a \emph{numerically exact} thermofield-based chain-mapping approach with MPS (TCMPS). The scheme evolves the total Hamiltonian $H_{\rm tot}$ that comprises the system and the bath. The TCMPS approach contains four main ingredients: (i) discretization of the bath with respect to its spectral density. (ii) Thermofield transformation that allows one to exactly map the effect of a finite-temperature bath to that of two zero-temperature baths. 
(iii) Star-to-chain mapping to ensure that the baths are mapped to linear chains. (iv) An MPS implementation to evolve the total Hamiltonian of the system plus baths. This approach was first introduced and rigorously tested in \cite{deVegaBanuls2015, deVegaSchollwock2015} and also used in \cite{GuoPoletti2017, CascioDeVega2018}. More details on the method and relevant convergence tests can be found in Appendix \ref{App:tcmps-error}. It should be pointed out, however, that the method is restricted to finite times due to the finiteness of the bath. Before the boundary is reached, the finite bath mimics an infinite reservoir, allowing us to compare it with our Redfield implementation. The results for the Redfield (solid lines) and the TCMPS (open squares) match exactly for the entire duration of the evolution considered herein, as seen in Fig.~\ref{fig:fig4}, validating our Redfield implementation and establishing its correctness over the results from the Lindblad master equations.

\section{\label{Sec:conclusion} Conclusions}
We have presented an implementation of the Redfield master equations using MPS and MPO. Unlike the conventional approach that requires the full eigenenergy spectrum, the MPS/MPO-based method allows us to probe the dynamics of large many-body open quantum systems. We have compared results from the Redfield master equation to typical master equations in Lindblad form which can be computed efficiently for large systems, and we have shown that those Lindblad master equations fail to capture the dynamics as the Redfield master equation can. Moreover, the time dependence in the evolution equations of our approach do not come at an additional cost and in most of the regimes it is computationally cheaper than the time-independent counterpart. The approach is thus robust, and the current algorithm can be readily extended to the study of multiple baths, different types of couplings, or even systems with time-dependent Hamiltonians. 

More work would be needed to increase the efficiency of the code, especially in terms of memory requirements, for example, using different evolution or integration schemes. 
Systematic comparison to the TCMPS approach, or to finite-time unitary evolution with small baths (see, e.g., \cite{MascarenhasSavona2017}), which are valid also for strong system-bath coupling, would give important insights into the regime of validity of the weak-coupling approximation \cite{ThingnaWangHanggi2012, ThingnaWangHanggi2013}.

The possibility of studying accurately the open dynamics of many-body quantum systems beyond Lindblad master equations leads to interesting opportunities in various directions, for instance, quantum thermodynamics and quantum transport.

\section*{\label{Sec:acknowledge} ACKNOWLEDGMENTS}
  D.P. acknowledges fruitful discussions with S. Maniscalco and K. Modi. D.P. and X.X. acknowledge support from the Ministry of Education of Singapore AcRF MOE Tier II (Project MOE2016-T2-1-065, WBS R-144-000-350-112). C.G. acknowledges support from the National Natural Science Foundation of China under Grants No. 11504430 and No. 11805279. J.T. acknowledges support from the European Research Council project NanoThermo (ERC-2015-CoG Agreement No. 681456). This research was supported in part by the National Science Foundation under Grant No. NSF PHY-1748958. The computational work for this article was partially performed on resources of the National Supercomputing Centre, Singapore \cite{nscc}.    

\appendix 
\setcounter{equation}{0}
\renewcommand{\theequation}{A\arabic{equation}} 

\section{\label{App:tcmps} Thermofield-based Chain-mapping Approach with Matrix Product States}
A detailed description of the {thermofield-based chain-mapping technique with matrix product states (TCMPS) to study many-body open quantum systems} can be found in Ref.~\cite{deVegaBanuls2015}.
Instead of studying the reduced dynamics of the system, the exact dynamics of the total composite system is investigated without perturbative treatments on the system-bath coupling strength. For our setup, we consider a {linearly discretized} bath with the frequency spacing given by $\Delta \omega = \omega_{\rm max} / N_{\rm max}$, where $\omega_{\rm max}$ is the numerical cutoff for the frequency of spectral density ${\rm J}(\omega)$ and $N_{\rm max}$ is the number of sites in the bath. The discretized bath Hamiltonian and system-bath coupling Hamiltonian are given by 
\begin{align}
	H_{\rm B}^{\rm discrete} & = \sum^{N_{\rm max}}_{j=1} \omega_{j} b^\dagger_j b_j, \\
	H_{\rm SB}^{\rm discrete}& = \sum^{N_{\rm max}}_{j=1}\sqrt{{\rm J}_j}\sigma^x_0 \left(b_j+b^\dagger_j\right),
\end{align}
where $\omega_j=j\Delta\omega$ and ${\rm J}_j = \int^{\omega_{j+1}} _{\omega_j} d\omega {\rm J}(\omega)\approx {\rm J}(\omega_j)\Delta \omega$.

Via thermofield transformation, the finite-temperature bath is mapped to another environment of $2N_{\rm max}$ modes $a_{1,j}$ and $a_{2,j}$, but in a vacuum state \cite{deVegaBanuls2015}. 
The transformed bath Hamiltonian and system-bath coupling then become 
\begin{align}
	H_{\rm B}^{\rm thermal}  =&  \sum^{N_{\rm max}}_{j=1} \omega_j\left(a^\dagger_{1,j}a_{1,j}-a^\dagger_{2,j}a_{2,j}\right), \\ 
	H_{\rm SB}^{\rm thermal} =& \sum^{N_{\rm max}}_{j=1} g_{1,j}\sigma^x_0\left(a_{1,j}+a^\dagger_{1,j}\right) \nonumber\\
	& + \sum^{N_{\rm max}}_{j=1}g_{2,j}\sigma^x_0\left(a_{2,j}+a^\dagger_{2,j}\right), 
\end{align}
where $g_{1,j} = \sqrt{{\rm J}_j[1+N(\omega_j)]}$ and $g_{2,j} = \sqrt{{\rm J}_j N(\omega_j)}$, with $N(\omega) = 1/[\exp{(\omega/T)}-1]$ for a harmonic oscillator bath. Since $\sigma^x_0$ is the operator that couples the system to the bath, we refer to it as the system operator.  	 

The above form is the so-called star configuration, where all modes of the bath are coupled to the system. {However, such a configuration can be computationally inefficient to evolve numerically}. The star-to-chain mapping \cite{Wilson1975, PriorPlenio2010, ChinPlenio2010, deVegaSchollwock2015} is then performed to transform the star configuration to a linear chain, which could be efficiently implemented with matrix product states, and which could be easier for a Trotter-expansion-based time evolution algorithm with matrix product states. The transformed Hamiltonians are  
\begin{align}
	H_{\rm B}^{\rm chain}  = & \sum_{k=1}^2 \sum^{N^\prime_{\rm max}}_{j=1} \Omega_{k,j} a^\dagger_{k,j}a_{k,j} \nonumber \\
	& +   \sum_{k=1}^2 \sum^{N^\prime_{\rm max}-1}_{j=1} \beta_{k,j} \left(a^\dagger_{k,j} a_{k,j+1} + a^\dagger_{k,j+1} a_{k,j}\right), \\
	H_{\rm SB}^{\rm chain} &  = \sum_{k=1}^2 \beta_{k,0}\sigma^x_0\left(a_{k,1}+a^\dagger_{k,1}\right), 
\end{align}
where we used $k$ to label the $N'_{\rm max} \le N_{\rm max}$ modes of two virtual baths $\Omega_{k,j}$, while the $\beta_{k,j}$ are generated via the Lanczos tridiagonalization of the discretized bath dispersion given by a diagonal matrix with elements, in increasing order, $\omega_1,\omega_2,\cdots,\omega_{N_{\rm max}}$. Here we emphasize that $N'_{\rm max}$ is the number of sites we kept in the transformed chain. The particular choice of $N'_{\rm max}$ could depend on the time scale of the simulation. 

In summary, the discretized bath Hamiltonian has undergone the following transformation for efficient simulations:
\begin{align}
  H_{\rm B}^{\rm discrete} {\xrightarrow[\text{transformation}]{\text{thermofield}}} H_{\rm B} ^{\text{thermal}} {\xrightarrow[\text{mapping}]{\text{star-to-chain}}} H_{\rm B}^{\rm chain}. \nonumber
\end{align}

In Fig. 4 of the main paper we use for the wave function bond dimension $D_f=100$, and at each site we consider a local Hilbert space of at most $d=5$ levels. The numerical simulation is done using a second-order Suzuki-Trotter method with time step $dt=0.01$.      
We find that for Fig. 4 we can take $\omega_{\rm max}=60$, $N_{\rm max}=6000$, $N'_{\rm max}=300$, to ensure that the results are converged in all relevant parameter regimes.

\section{\label{App:error} Numerical Error}
\subsection{\label{App:rmemps-error} Redfield Master Equation with MPS and MPO}
\begin{figure}[t]
	\centering
        \includegraphics[width=\columnwidth]{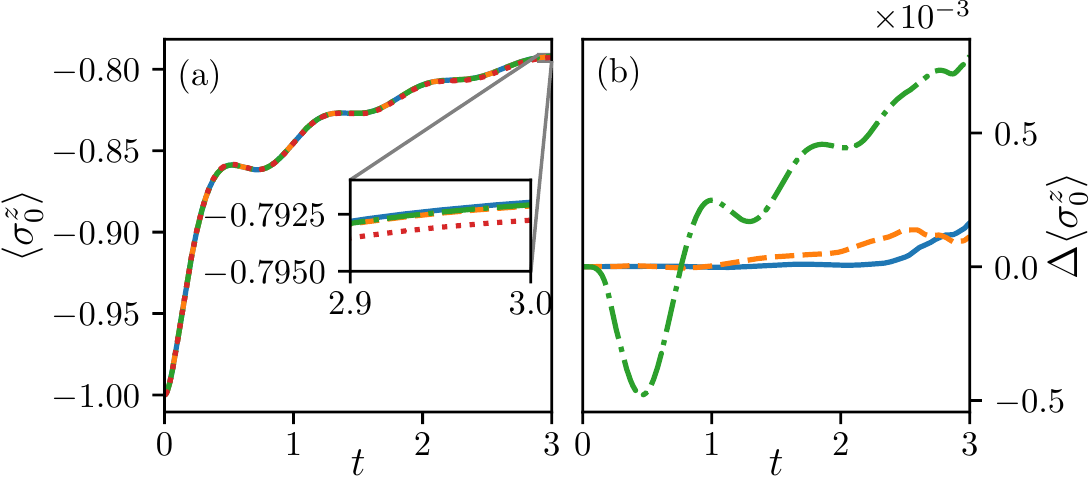}
	\caption{(a) The dynamics of local magnetization of the central site $\Braket{\sigma_0^z}$ under various system operator bond dimensions $D_W=15$ (red dotted), $30$ (green dash-dotted), $45$ (orange dashed), and $60$ (blue solid). (b) The difference of the local magnetization for between bond dimension $D_W=60$ and $15$ (green dash-dotted), $60$ and $30$ (orange dashed), $60$ and $45$ (blue solid). The bond dimension of the density operator is $D=100$. Other parameters are consistent with Fig. 2 in the article.}
	\label{fig:S1}
\end{figure}

Our numerical simulation relies on the truncation of the evolution of the system operator $\sigma^x_0$ as well as the density operator $\rho$. We perform the following error analysis by varying the bond dimension of the system operator and the bond dimension of the density operator. We first investigate the error due to truncation of the system operator. By using the parameters in Fig. 2, we check the results for the system operator with bond dimensions $D_W=15,\;30, \;45$, and $60$. From Fig. {\ref{fig:S1}}(a), the dynamics show qualitative agreement for various bond dimensions. In Fig. {\ref{fig:S1}}(b), we show the differences in results from different bond dimensions $D_W$ (i.e., between $60$ and $15$, $60$ and $30$, $60$ and $45$) . By keeping system operator bond dimension as $D_W=30$, the error of our results would be of the order of $10^{-4}$.

\begin{figure}[t]
	\centering
        \includegraphics[width=\columnwidth]{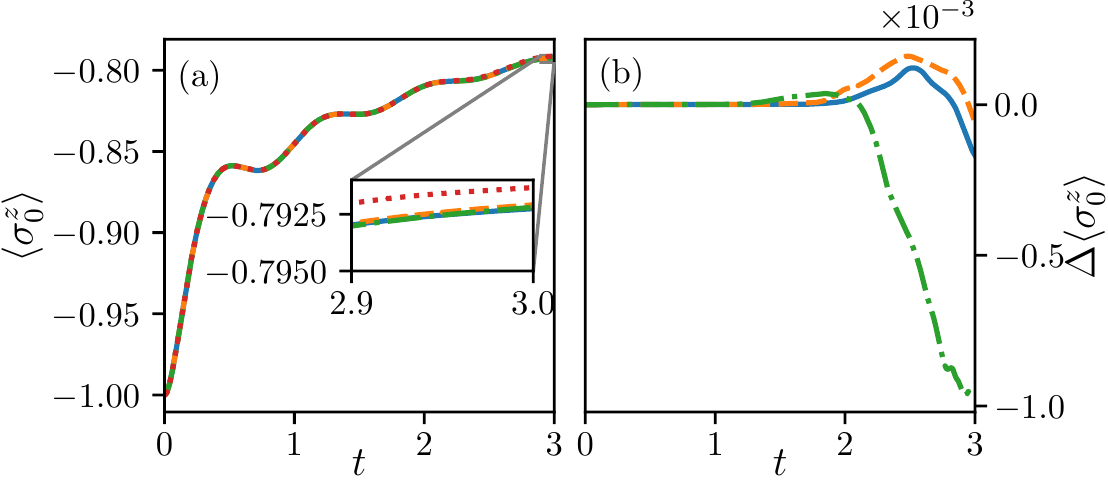}
	\caption{(a) The dynamics of local magnetization of the central site $\Braket{\sigma_0^z}$ under various density operator bond dimensions $D=50$ (red dotted), $75$ (green dash-dotted), $100$ (orange dashed), $125$ (blue solid). (b) The difference of the local magnetization for between bond dimension $D=125$ and $50$ (green dash-dotted), $125$ and $75$ (orange dashed), $125$ and $100$ (blue solid). The bond dimension of the system operator is $D_W=30$. Other parameters are consistent with Fig. 2 in the article.}
	\label{fig:S2}
\end{figure}

For various bond dimensions $D$ of the density operator, the results are shown in Fig. \ref{fig:S2}. The dynamics obtained via various density operator bond dimensions again show agreement with each other. The error is of the order of $10^{-4}$ when the density operator bond dimension is kept at $D=100$.

\subsection{\label{App:tcmps-error} Thermofield-based Chain-mapping Approach with MPS}
In this section, we show the error for the TCMPS approach. Many parameters can be fine-tuned, for instance, the discretization of the spectral function, {the numerical frequency cutoff}, the Trotter evolution parameters, the size of the local Hilbert space, and the maximum bond dimension. Here we focus on the discretization of the bath and on the bond dimension. For the discretization of the bath, we first check the error with respect to the bath discretization parameters $\Delta \omega$ and $\omega_{\rm max}$.

\begin{figure}[t]
	\centering
	\includegraphics[width=\columnwidth]{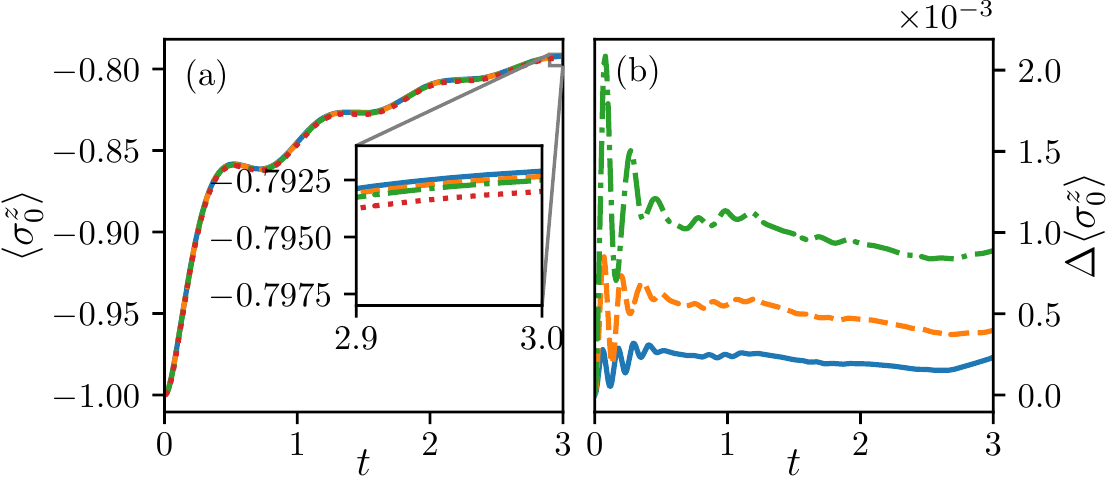}
        \caption{(a) The dynamics of local magnetization of the central site $\Braket{\sigma_0^z}$ under various numerical cutoff frequencies $\omega_{\rm max}=50$ (red dotted), 60 (green dash-dotted), 70 (orange dashed), 80 (blue solid). (b) The difference of the local magnetization between various numerical cutoff frequencies $\omega_{\rm max}$ 80 and 50 (green dash-dotted), 80 and 60 (orange dashed), 80 and 70 (blue solid). The bath discretization parameters are $\Delta\omega = 0.01$. The transformed chain size $N'_{\rm max}=300$ with local bath dimension $d=5$. The wave function bond dimension $D_f=100$. Other parameters are consistent with Fig. 2 in the article.}
	\label{fig:S3}
\end{figure}

By comparing the results of various numerical frequency cutoffs $\omega_{\rm max}=50$, $60$, $70$, $80$, it can be shown that the dynamics reach a good agreement in Fig. \ref{fig:S3}(a). A large numerical cutoff frequency $\omega_{\rm max} = 60$ is required due to the shape of the spectral function to obtain an error at the order of $10^{-4}$ as illustrated by Fig. \ref{fig:S3}(b).  

\begin{figure}[t]
	\centering
	\includegraphics[width=\columnwidth]{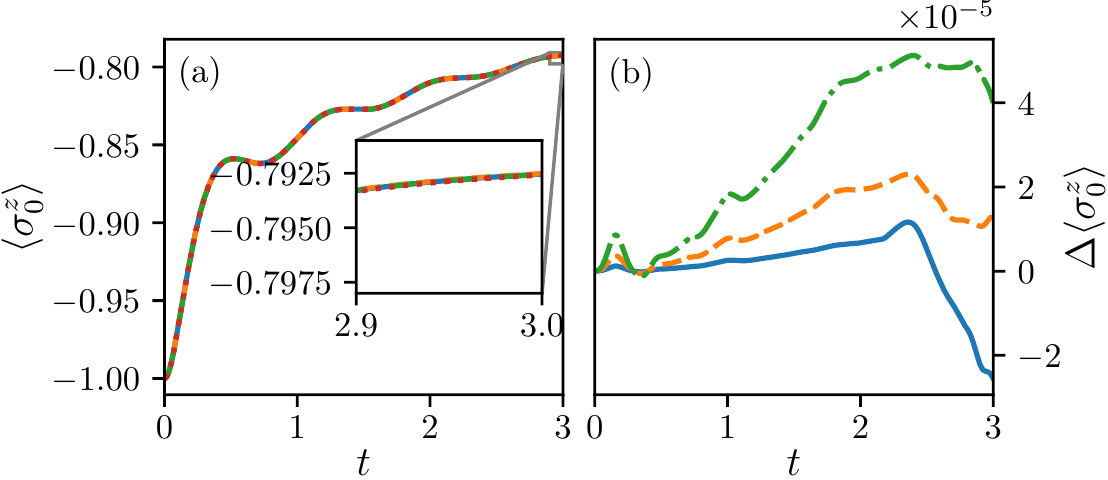}
        \caption{(a) The dynamics of local magnetization of the central site $\Braket{\sigma_0^z}$ under various frequency spacings $\Delta \omega$ = 0.04 (red dotted), 0.02 (green dash-dotted), 0.01 (orange dashed), 0.005 (blue solid). (b) The difference of the local magnetization between various frequency spacings, $\Delta \omega$, 0.005 and 0.04 (green dash-dotted), 0.005 and 0.02 (orange dashed), 0.005 and 0.01 (blue solid). The bath numerical cutoff frequency $\omega_{\rm max}=60$ and the transformed chain size $N'_{\rm max}$ are kept at 300 with local bath dimension $d=5$. The wave function bond dimension $D_f$=100. Other parameters are consistent with Fig. 2 in the article.}
	\label{fig:S4}
\end{figure}

It follows that we would also need to consider the frequency spacing $\Delta \omega = 0.005$, $0.01$, $0.02$, $0.04$, which determines the $N_{\rm max}$ ranging from $12000$ to $1500$ with a fixed $\omega_{\rm max}=60$. By studying the dynamics in Fig. \ref{fig:S4}(a), the dynamics for frequency spacing $\Delta \omega$ cannot be resolved at $10^{-3}$. The errors are more quantitatively depicted in Fig. \ref{fig:S4}(b), where clear convergence can be observed when the frequency spacings are reduced. We have represented the difference between various bath sizes. In particular. Fig. \ref{fig:S4}(b) shows the difference $\Delta\langle\sigma^z_0\rangle$ between frequency spacings $\Delta\omega=0.005$ and $0.04$ (green dash-dotted), $0.005$ and $0.02$ (orange dashed), $0.005$ and $0.01$ (blue solid). In our simulation, by choosing $\Delta \omega=0.01$, the error would be at the order of $10^{-5}$.

\begin{figure}[t]
	\centering
	\includegraphics[width=\columnwidth]{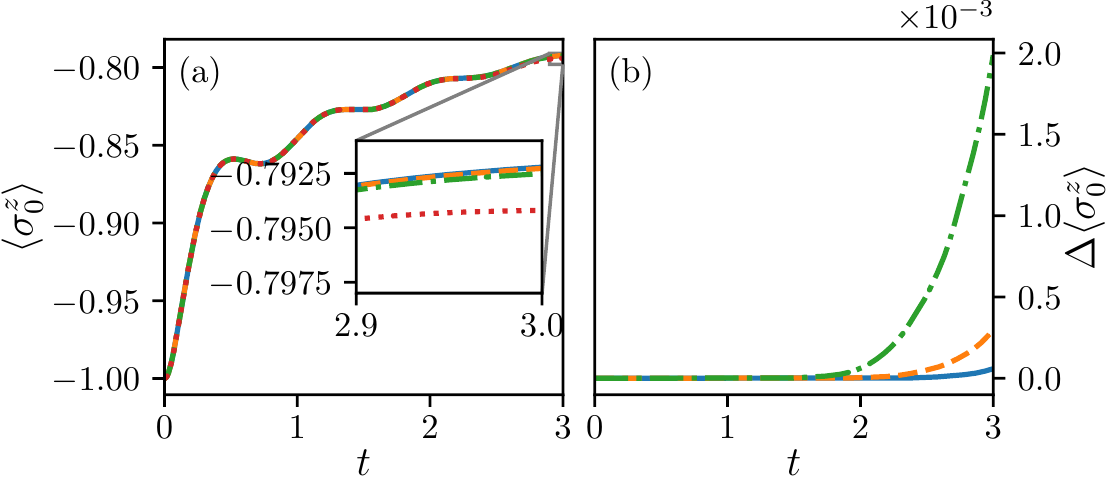}
        \caption{(a) The dynamics of local magnetization of the central site $\Braket{\sigma_0^z}$ under various wave function bond dimensions $D_f=50$ (red dotted), 100 (green dash-dotted), 150 (orange dashed), 200 (blue solid). (b) The difference of the local magnetization between bond dimensions $D_f=200$ and 50 (green dash-dotted) , 200 and 100 (orange dashed), 200 and 150 (blue solid). The bath discretization parameters are $N^\prime_{\rm max}=300$, $\Delta\omega = 0.01$, and $\omega_{\rm max} = 60$ with local bath dimension $d=5$. Other parameters are consistent with Fig. 2 in the article.}
	\label{fig:S5}
\end{figure}

Last we examine the role of the wave function bond dimension $D_f$ kept, ranging from $50$ to $200$. Fig. \ref{fig:S5}(a) also reveals qualitative agreement between various bond dimensions, while Fig. \ref{fig:S5}(b) demonstrates the error convergence when the bond dimension increases.

In the above simulation, we used a second-order Trotter time evolution with a time step of $0.01$, resulting in a relatively large error. This could be improved with a fourth-order Trotter method. However, increasing the accuracy of the method, or its time of validity, could require a sizable computing time.

\section{\label{App:rme} Comparison between Redfield Master Equation and Time-independent Redfield Master Equation}

\begin{figure}[t]
	\centering
	\includegraphics[width=\columnwidth]{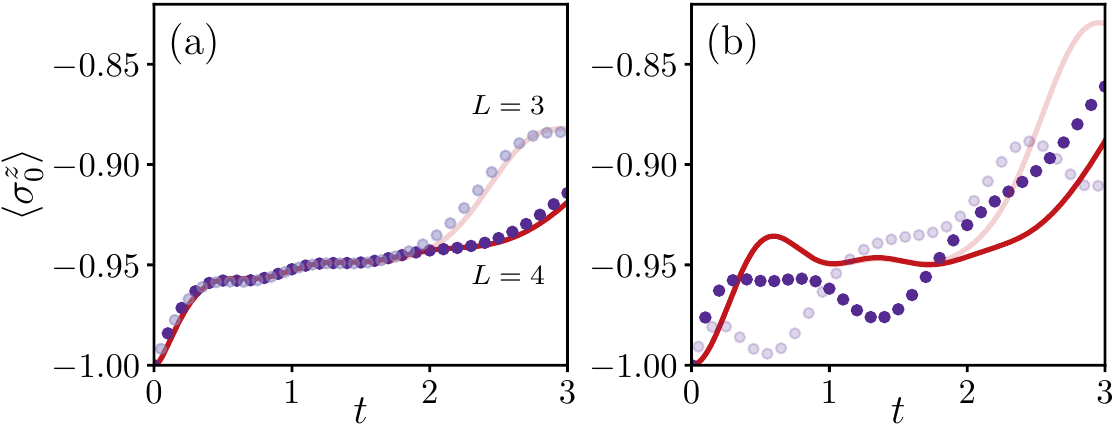}
	\caption{The dynamics of local magnetization of the central site $\Braket{\sigma_0^z}$ vs time $t$ computed from RME (red solid lines) and iRME (purple filled circles) for (a) $\Delta$ = 0.5, $T=2$, $\omega_{\rm c}$=10, $\gamma$=0.02 and (b) $\Delta$ =0.5, $T=2$, $\omega_{\rm c}$=1, $\gamma=0.2$ for seven sites (faint colors) and nine sites (darker colors)}
	\label{fig:S6}
\end{figure}

It is also important to point out that in the small $\omega_{\rm c}$ regime, the evolution due to the time-dependent RME cannot be approximated by the (time-independent) iRME. This is highlighted in Figs.~\ref{fig:S6}(a) and \ref{fig:S6}(b), where results from RME (red solid lines) are compared to those of its time-independent approximation iRME, where $\mathcal{R}^{\infty}$ is used instead of $\mathcal{R}^{t}$ (purple circles). Here we consider systems with nine (darker lines or circles) or seven (lighter lines or circles) sites. In Fig.~\ref{fig:S6}(a) we consider $\omega_{\rm c}=10$ and in Fig.~\ref{fig:S6}(b) $\omega_{\rm c}=1$. For a large enough cutoff, panel (a), the predictions of RME and iRME are in agreement. For small cutoffs $\omega_{\rm c}$, panel (b), the finite time effects are stronger and the inaccuracy of the iRME more evident. This is due to the fact that at small frequencies the size of the system plays a bigger role. For the iRME the superoperator $\mathcal{R}^\infty$ would be quickly affected by the finite system size, while $\mathcal{R}^t$ would require some time before the finite-size effects were felt.



\begin{thebibliography}{10}
\bibitem{details} In all our simulations we use maximum bond dimensions $D=100$ and $D_W=30$. The time step in the second order Trotter evolution of the MPO is 0.00125, and in the fourth-order Runge-Kutta evolution of the density matrix, it is 0.01. For the Romberg integration we use a third order with time step $0.00125$. For the evolution of Lindblad master equations we use $D=100$ with fourth-order Trotter-Suzuki method with time step 0.01.
\bibitem{hfield} We note here that the presence of a local magnetization $h$ is a prerequisite for LLME to perform a local approximation, while it is not necessary for RME.
\bibitem{spectral} From the definition of $C(\tau)$ we get $C(\tau)=\int_0^\infty \left(d\omega/\pi\right){\rm J}(\omega) \left[\coth\left(\omega/2T\right)\cos(\omega\tau) - \ii \sin(\omega\tau)\right]$ with ${\rm J}(\omega)=\pi \sum c_n^2/(2m_n\omega_n)\delta(\omega-\omega_n)$.
\end{thebibliography}
\end{document}